\documentclass{article}[12pt]
\usepackage{geometry}[left=2in, right=2in, top=1.25in, bottom=1.25in]
\usepackage{setspace}
\usepackage{graphicx} 
\usepackage{amsmath, amssymb}
    \allowdisplaybreaks
\usepackage{amsthm}
    \newtheoremstyle{customstyle}
      {1.5\topsep} 
      {1.5\topsep} 
      {\itshape} 
      {} 
      {\bfseries \itshape} 
      {.} 
      {.5em} 
      {} 
    \theoremstyle{customstyle}
        
        \newtheorem*{theorem*}{Theorem} 
        \newtheorem{definition}{Definition}
        \newtheorem{proposition}{Proposition}
        
        \newtheorem{assumption}{Assumption}
        
\usepackage{tikz}
\usepackage{pgfplots}
    \pgfplotsset{compat=1.16}
\usepackage{appendix}
\usepackage{enumitem}
    \setlist[enumerate]{topsep=2pt, itemsep=2pt, parsep=0pt}
\usepackage{xcolor}
\usepackage{natbib}
\usepackage{authblk}
\usepackage[ruled,vlined]{algorithm2e}
    \SetArgSty{textnormal}

\title{
Transferable Utility Matching Beyond Logit:\\
Computation and Estimation with General Heterogeneity
}

\author[1]{A.\ Galichon\thanks{Economics Department, FAS, and Mathematics Department, Courant Institute, New York University; and Economics Department, Sciences Po. Email: ag133@nyu.edu. Galichon gratefully acknowledges funding from ERC grant CoG-866274.},
A.\ Jacquet\thanks{Economics Department, Sciences Po. Email: antoine.jacquet@sciencespo.fr. Funding from ERC grant CoG-866274 is gratefully acknowledged.},
G.\ Salakhutdinov\thanks{Ecole Polytechnique. Email: georgy.salakhutdinov@polytechnique.edu.}
}

\date{\today}

\begin{document}

\vspace{-150pt}

\maketitle

\begin{abstract}
We present a general framework for matching with transferable utility (TU) that accommodates arbitrary heterogeneity without relying on the logit structure.
The optimal assignment problem is characterized by tractable linear programming formulation, allowing flexible error distributions and correlation patterns.
We introduce an iterative algorithm that solves large-scale assignment problems with guaranteed convergence and an intuitive economic interpretation, and we show how the same structure supports a simulated moment-matching estimator of the systematic surplus.
Experiments using simulated data demonstrate the algorithm’s scalability and the estimator’s consistency under correct specification, as well as systematic bias arising from logit misspecification.
\end{abstract}

\newpage


\section{Introduction}

Empirical work on matching has reshaped how economists measure sorting, evaluate policy, and interpret the formation of two-sided relationships such as marriages, employment, and college admissions.
The model by \citet{ChooSiow2006}, in particular, has become a popular framework to analyze markets under the transferable utility (TU) assumption. 
The appeal of this framework lies in large part in its analytical tractability: when the agents' idiosyncratic preference shocks are assumed to be i.i.d.\ extreme value type-I (Gumbel), equilibrium matching patterns admit a closed-form logit structure that maps directly from surplus parameters to observed match frequencies.
This assumption has fostered a vast empirical literature on assortative matching by education, income, age, ethnicity, and other attributes \citep[e.g.,][]{ChiapporiSalanieWeiss2017, GousseJacquemetRobin2017, BisinTura2019, Ciscato2025}.

Yet the same assumption that ensures tractability can also be a serious limitation.
In modern datasets of the marriage market for instance, types can usually be defined as the intersection of several observable attributes since large administrative data now allow rich cross-classifications.
As the number of attributes grows, it becomes increasingly implausible that an individual's unobserved preferences for all composite options are independent.
For instance, a person's idiosyncratic value for a highly educated urban partner is likely to be correlated with her value for a highly educated rural partner, since both options share the same education component.
When independence is imposed by construction, substitution patterns and counterfactual responses become mechanically constrained, and estimates of sorting strength or welfare effects may be distorted.

This paper develops a general and computationally tractable framework that relaxes the i.i.d.\ Gumbel assumption while retaining the economic structure that underlies the original Choo--Siow model.
Our framework allows for arbitrary distributions of the idiosyncratic preference shocks, including rich correlation structures induced by multi-attribute types.
For example, we can let the shock for a composite type $y=(y_1,y_2)$ decompose additively across its attributes, $\varepsilon_{i,y}=\varepsilon_{i,y_1}+\varepsilon_{i,y_2}$, where the attribute-level shocks are independent.
This simple construction introduces correlation across composite options that share such attributes and encompasses a wide class of probit and nested-logit specifications without committing to any closed-form functional form.

Our contribution is essentially twofold.
In a first step, we propose an algorithm to solve in practice large-scale optimal assignment problems à la \citet{Shapley-Shubik1971}.
This method is grounded in tools developed to solve large-scale linear programs, specifically the Dantzig--Wolfe decomposition \citep{DantzigWolfe1960}.
We discuss the natural economic interpretation of our algorithm as a series of discrete choice problems, and we use numerical simulations to demonstrate its performance in solving large optimal assignment problems.
Depending on the scale of the problem to solve, we record up to 25 times computational gains for our algorithm compared to a state-of-the-art general-purpose solver.

In a second step, we build on this formulation to provide a method for estimating the matching surplus, which is suited to large-scale data and general distributions.
Following \citet{GalichonSalanie2022}, we recover (a parametrized version of) the systematic surplus matrix $\Phi$ from observed type-by-type match frequencies using a moment-matching approach.
When the idiosyncratic shocks are i.i.d.\ Gumbel, estimation can be performed using closed-form formulas.
In general however it cannot, and we therefore rely on a method by simulation.
We show that the problem of estimating the matching surplus is then equivalent to a finite assignment problem with simulated agents, which we can solve using our algorithm discussed above.
Again using numerical experiments, we demonstrate the consistency of our estimator on an example with probit errors, and exhibit a consistent bias when using a misspecified error distribution.

Our approach relates to several strands of the literature.
Within the TU matching tradition, our methodology builds directly on \citet{ChooSiow2006} and \citet{GalichonSalanie2022}, maintaining separability but dispensing with the distributional restrictions that yield logit formulas.
In spirit, our treatment of correlated unobserved heterogeneity parallels advances in single-agent discrete-choice models, where probit and nested-logit specifications generalized the logit without sacrificing interpretability \citep{McFadden1981, Train2009}.
Our estimator also takes inspiration from the method of simulated moments developed by \citet{McFadden1989}.
Existing works have generalized the Choo--Siow model in other dimensions.
\citet{GalichonSalanie2022} analyze the TU model of Choo and Siow under general distributional assumptions of the heterogeneity, although they do not explore the practical computational aspects studied here.
\citet{GualdaniSinha2023}, coincidentally also using techniques from linear programming, investigate partial identification of the matching surplus when the distributions of the preference heterogeneity is not known.
\citet*{ChiapporiNguyenSalanie2019} investigate the bias which can result from mistakenly imposing the separability assumption.

The paper proceeds as follows.
Section \ref{sec:model} presents the model and motivates the role of separability under general error distributions.
Section \ref{sec:optimal_assignment} introduces the type-aggregated formulation and the RROA algorithm, establishes its convergence, and documents its computational performance.
Section \ref{sec:estimation} extends the analysis to estimation, defining the simulated social surplus and showing how the same algorithmic structure delivers an efficient simulated moment-matching estimator.
Together, these results demonstrate that credible empirical matching with many attributes need not rely on the i.i.d.\ Gumbel assumption: flexible heterogeneity and large-scale data are now compatible within the transferable utility framework.
Proofs for our formal statements can be found in appendix \ref{apx:proofs}.


\section{Model}
\label{sec:model}

We study a bipartite, one-to-one matching market with transferable utility à la \citet{ChooSiow2006}.
Matching is static, frictionless, and individuals have complete information on potential partners' types.
The crucial difference between our framework and the one from \citet{ChooSiow2006} concerns the distribution of the individual heterogeneity: While they assume that shocks are i.i.d.\ Gumbel, we remain agnostic regarding their distribution.
In this respect, our framework is closest to \citet{GalichonSalanie2022}.

\paragraph{Matching.}
Consider a population of women $i \in I$ and men $j \in J$, where $I$ and $J$ are finite sets.
Each woman or man belongs to a type $x \in X$ or $y \in Y$, respectively;
occasionally we may denote $x_i$ the type of woman $i$, and $y_j$ the type of man $j$.
Because the population is finite, the number of observed types must also be finite.
But beyond that, we think about the type sets $X$ and $Y$ as being orders of magnitude smaller than the populations $I$ and $J$.
This reflects two facts.
First, in empirical practice, types are defined as intersections of observed characteristics, which typically yield fewer types than individuals.
Second, types are kept intentionally coarse in order to preserve statistical power.

A \emph{matching} specifies who matches with whom.
In our finite-population framework, it is simply a matrix $\tilde \pi = (\tilde \pi_{ij})$ of size $|I \times J|$ with non-negative entries such that
\begin{align}
\textstyle\sum_j \tilde \pi_{ij} &\leq 1 \qquad \forall i \label{eq:indiv_constraint_i} \\
\textstyle\sum_i \tilde \pi_{ij} &\leq 1 \qquad \forall j \label{eq:indiv_constraint_j}.
\end{align}
Conditions \eqref{eq:indiv_constraint_i}--\eqref{eq:indiv_constraint_j} have different interpretations depending on whether the matching is pure or fractional.
A matching $\pi$ is called \emph{pure} when $\tilde \pi_{ij} \in \{0,1\}$ for all $ij$.
In this special case, we can interpret each $\tilde \pi_{ij}$ as the indicator that $i$ and $j$ are matched, hence conditions \eqref{eq:indiv_constraint_i}--\eqref{eq:indiv_constraint_j} mean that any individual should be matched to at most one partner.
Individual-level data on married couples, for instance, typically involves pure matchings.
When a matching is not pure however, it is called \emph{fractional}.
When fractional matchings are allowed, $\tilde \pi_{ij}$ can be interpreted instead as the fraction of time that $i$ is matched with $j$, and conditions \eqref{eq:indiv_constraint_i}--\eqref{eq:indiv_constraint_j} mean that any individual has a single unit of time to dispense across partners.
Such data could occur for instance on labor markets with part-time workers.
Even though our framework accommodates both pure and fractional matchings, we will often speak as if $\tilde \pi_{ij}$ were the indicator that $i$ and $j$ are matched to keep exposition simple.

\paragraph{Surplus and individual heterogeneity.}
We now turn to the value created by a match.
When woman $i$ and man $j$ form a pair, they generate a joint economic value $\tilde \Phi_{ij}$.
Following \citet{ChooSiow2006}, we assume that this joint value is \emph{separable} in the following sense.

\begin{assumption}[Separability] \label{ass:separability}
There exists a matrix $\Phi = (\Phi_{xy})$ such that:
\begin{enumerate}[label=(\roman*)]
\item the joint value of a match between woman $i$ of type $x$ and man $j$ of type $y$ is
\vspace{-6pt}
\begin{equation} \label{eq:separability}
\tilde \Phi_{ij} = \Phi_{xy} + \varepsilon_{iy} + \eta_{xj},
\vspace{-6pt}
\end{equation}
\item the singlehood utility of woman $i$ is $\varepsilon_{i0}$,
\item the singlehood utility of man $j$ is $\varepsilon_{0j}$,
\end{enumerate}
where, conditional on $x_i = x$, the $(|Y|+1)$-dimensional random vector $\varepsilon_i = (\varepsilon_{iy})_{y \in Y \cup \{0\}}$ has probability distribution $\mathbf P_x$, and, conditional on $y_j = y$, the $(|X|+1)$-dimensional random
vector $\eta_j = (\eta_{xj})_{x \in X \cup \{0\}}$ has probability distribution $\mathbf Q_y$.
\end{assumption}

According to Assumption \ref{ass:separability}, the joint value $\tilde \Phi_{ij}$ is the sum of three terms:
a systematic part $\Phi_{xy}$, which only depends on the type pair $xy$ of the matched agents;
and two idiosyncratic preference shocks $\varepsilon_{iy}$ and $\eta_{xj}$ of the agents over potential partner types.
The systematic part $\Phi_{xy}$ is called the \emph{systematic surplus} of the match.%
\footnote{The match surplus is
$\tilde \Phi_{ij} - \varepsilon_{i0} - \eta_{0j} = \Phi_{xy} + (\varepsilon_{iy} - \varepsilon_{i0}) + (\eta_{xj} - \eta_{0j})$,
so the name systematic surplus for $\Phi_{xy}$ is warranted as long as $\mathbf E [\varepsilon_{iy} - \varepsilon_{i0}] = 0$ and $\mathbf E [\eta_{xj} - \eta_{0j}] = 0$.
This is however without loss of generality, since we can redefine
$\varepsilon_{iy}' = \varepsilon_{iy} - \mathbf E [\varepsilon_{iy} - \varepsilon_{i0}]$, 
$\eta_{xj}' = \eta_{xj} - \mathbf E [\eta_{xj} - \eta_{0j}]$, 
and $\Phi_{xy}' = \Phi_{xy} + \mathbf E [\varepsilon_{iy} - \varepsilon_{i0}] + \mathbf E [\eta_{xj} - \eta_{0j}]$.
}
The separability assumption notably entails that individuals are indifferent between all potential partners of the same type.
It is crucial to the identification of the systematic surplus $\Phi_{xy}$ \citep{GalichonSalanie2022}.

\citet{ChooSiow2006} go further than Assumption \ref{ass:separability} since they assume that the error terms $\varepsilon_{iy}, \varepsilon_{i0}, \eta_{xj}, \eta_{0j}$ are i.i.d.\ Gumbel (extreme value type I), leading to convenient closed-form estimates of the systematic surplus.
We wish to generalize their approach, and as such we do not make any ex ante assumption on the distribution of these error terms for most of our analysis.
To see why considering such general distributions may be important, consider the following example where correlation in the idiosyncratic preferences arises naturally from the structure of the type sets $X$ and $Y$.

\paragraph{Example.}
Consider a marriage market where each agent's observable type $x \in X$ (and symmetrically $y \in Y$) is defined by two characteristics, $x = (x_1, x_2)$.
For instance, $x_1$ could be the education level (whether the agent has a college diploma), and $x_2$ the region of origin (whether the individual is from a rural or urban background).
In the standard Choo--Siow framework, the idiosyncratic preference for, say, a highly educated urban woman is assumed to be independent of that for a highly educated rural woman.
Now consider a more structured alternative, where the idiosyncratic shock attached to the composite type $x = (x_1, x_2)$ is the sum of two independent components drawn at the level of each attribute:
\[
\eta_{xj} = \eta_{x_1, j} + \eta_{x_2, j}.
\]
This natural formulation induces correlation in idiosyncratic preferences across composite types sharing common characteristics.
This captures, for instance, that an agent's preference over partners' education levels may be systematically related across regions of origin.

\paragraph{}
In the spirit of the previous example, general distributions of the idiosyncratic preferences allow us to consider a wide range of models beyond the i.i.d.\ Gumbel case, such as probit models with arbitrary covariance matrices or nested logit models.

\section{Optimal assignment}
\label{sec:optimal_assignment}

In this section we study the optimal assignment problem and its computation when the population size becomes very large.
The computational tools that we introduce here lay the groundwork for the estimation method that we present in section \ref{sec:estimation}.

Given realized match values $\tilde \Phi_{ij}$ and singlehood utilities $\varepsilon_{i0}$ and $\eta_{0j}$, the optimal assignment problem consists in finding a matching $\tilde \pi$ which maximizes the total surplus in the population:
\begin{align}
\label{eq:optimal_matching_naive}
\max_{\tilde \pi_{ij} \geq 0} ~ & \sum_{ij} \tilde \pi_{ij} (\tilde \Phi_{ij} - \varepsilon_{i0} - \eta_{0j}) \tag{$\tilde{\mathcal A}$} \\
\text{s.t.} ~ & \textstyle\sum_j \tilde \pi_{ij} \leq 1 \quad (\forall i)  \notag \\
              & \textstyle\sum_i \tilde \pi_{ij} \leq 1 \quad (\forall j). \notag
\end{align}
We call such a matching $\tilde \pi$ an \emph{optimal matching}.
It is well known since \citet{Shapley-Shubik1971} that any optimal matching can be decentralized as the equilibrium of a matching problem with transferable utility (TU), whereby the value $\tilde \Phi_{ij}$ created by a match $ij$ is split additively between the two partners.
Crucially, the resulting utilities $u_i$ and $v_j$ are recovered in the optimal assignment problem \eqref{eq:optimal_matching_naive} as the Lagrange multipliers of the constraints indexed by $i$ and $j$ respectively.
The TU assumption thus means that $u_i + v_j = \tilde \Phi_{ij}$ as soon as $\tilde \pi_{ij} > 0$.

We are interested in solving the optimal assignment problem \eqref{eq:optimal_matching_naive} when the size of the population becomes large.
This is a linear program with $|I \times J|$ variables and $|I + J|$ constraints, hence its size $|I||J| \times |I + J|$ is cubic in the population size.
If, for instance, we consider a typical dataset as having around 10,000 individuals for $I$ and $J$, the problem's size is of order $10^{12}$, rendering it intractable for standard linear solvers.
We therefore need an alternative method to tackle the problem.

\subsection{Reformulation under separability}

As a first step towards solving \eqref{eq:optimal_matching_naive}, we show that we can leverage the separability assumption in order to reduce the problem's size.
We introduce some notations: let $\delta_{ix} = \mathbf 1(x_i = x)$ and $\delta_{jy} = \mathbf 1(y_j = y)$ be indicators of individuals' type, and define
\begin{equation} \label{eq:values_match}
\alpha_{iy} = \sum_{x \in X} \delta_{ix} \frac{\Phi_{xy}}{2} + \varepsilon_{iy},
\qquad
\gamma_{xj} = \sum_{y \in Y} \delta_{jy} \frac{\Phi_{xy}}{2} + \eta_{xj}.
\end{equation}
The value $\alpha_{iy}$ is the utility obtained by woman $i$ when she matches with a man of type $y$, under the assumption that systematic surplus is split equally between partners.%
\footnote{This equal splitting the surplus by default is an arbitrary convention chosen to maintain symmetry: how surplus is actually divided between partners does not matter for the purpose of maximizing the total surplus.}
Similarly, $\gamma_{xj}$ is the utility obtained by man $j$ when he matches with a woman of type $x$.

Notice that, under Assumption 1, individuals are indifferent between potential partners of the same type, and therefore a matching need not keep track of exactly which $i$ matches with which $j$, but only of which \emph{type} each individual is matched with.
This leads us to introduce the new aggregated variables
\begin{equation} \label{eq:aggregate_indicators}
\pi_{iy} = \sum_j \delta_{jy} \tilde \pi_{ij},
\qquad
\pi_{xj} = \sum_i \delta_{ix} \tilde \pi_{ij},
\end{equation}
indicating respectively whether woman $i$ is matched with a man $y$, and whether man $j$ is matched with a woman $x$.
Similarly, we introduce the singlehood indicators
\begin{equation} \label{eq:singlehood_indicators}
\pi_{i0} = 1 - \sum_j \tilde \pi_{ij},
\qquad
\pi_{0j} = 1 - \sum_i \tilde \pi_{ij},
\end{equation}
indicating respectively whether woman $i$ and man $j$ are unmatched.
Using these aggregated variables, the optimal assignment problem \eqref{eq:optimal_matching_naive} admits the alternative formulation
\begin{align} \label{eq:optimal_matching_refined}
\max_{\pi_{iy}, \pi_{i0}, \pi_{xj}, \pi_{0j} \geq 0} ~ & \sum_i \bigg[ \pi_{i0} \varepsilon_{i0} + \sum_{y \in Y} \pi_{iy} \alpha_{iy} \bigg]
+ \sum_j \bigg[ \pi_{0j} \eta_{0j} + \sum_{x \in X} \pi_{xj} \gamma_{xj} \bigg]
\tag{$\mathcal A$} \\
\text{s.t.} ~ & \pi_{i0} + \textstyle\sum_{y \in Y} \pi_{iy} = 1 \qquad (\forall i) \notag \\
              & \pi_{0j} + \textstyle\sum_{x \in X} \pi_{xj} = 1 \qquad (\forall j) \notag \\
              & \textstyle\sum_i \delta_{ix} \pi_{iy} = \textstyle\sum_j \delta_{jy} \pi_{xj}  \quad (\forall xy) \notag.
\end{align}
Problem \eqref{eq:optimal_matching_refined} includes individual feasibility constraints indexed by $i$ and $j$, similar to those in \eqref{eq:optimal_matching_naive}.
Their Lagrange multipliers still correspond to the utilities $u_i$ or $v_j$ obtained by each individual.
The novelty lies in the balance condition $\sum_i \delta_{ix} \pi_{iy} = \sum_j \delta_{jy} \pi_{xj}$, which requires there to be as many women of type $x$ matched with men of type $y$, as there are men $y$ matched with women $x$.
This new constraint is a simple consequence of type aggregation.
Interestingly, the Lagrange multiplier associated with this constraint, which we denote $T_{xy}$, has a natural interpretation as the systematic transfer from women to men in matches $xy$.
This transfer is such that if a woman $i$ of type $x$ and a man $j$ of type $y$ are matched, then their respective utilities are
\begin{equation} \label{eq:utilities_match}
u_i = \frac{\Phi_{xy}}{2} - T_{xy} + \varepsilon_{iy}
\qquad \text{and} \qquad
v_j = \frac{\Phi_{xy}}{2} + T_{xy} + \eta_{xj}.
\end{equation}

The equivalence between the two problems \eqref{eq:optimal_matching_naive} and \eqref{eq:optimal_matching_refined} is made precise by the following result.

\begin{proposition} \label{prop:optimal_assignment_reformulation}
Under Assumption \ref{ass:separability},
\begin{itemize}[itemsep=0pt, topsep=4pt]
\item If $\tilde \pi$ is a solution to \eqref{eq:optimal_matching_naive}, then $\pi$ defined by \eqref{eq:aggregate_indicators}--\eqref{eq:singlehood_indicators} is a solution to \eqref{eq:optimal_matching_refined}.
\item If $\pi$ is a solution to \eqref{eq:optimal_matching_refined}, then any $\tilde \pi$ such that \eqref{eq:aggregate_indicators} holds for all $iy$ and $xj$ is a solution to \eqref{eq:optimal_matching_naive}.
In addition, such a $\tilde \pi$ always exists.
\end{itemize}
\end{proposition}

In the following, we focus on the formulation \eqref{eq:optimal_matching_refined}, which we also refer to as the optimal assignment problem.
Its solutions are also called optimal matchings.

Aside from questions of interpretation, the main advantage of formulation \eqref{eq:optimal_matching_refined} compared to \eqref{eq:optimal_matching_naive} is that its size has been reduced by one order of magnitude: this is now a linear program with $|I | |Y + 1| + |J| |X + 1|$ variables and $|XY| + |I + J|$ constraints, for a total size which is quadratic in the number of individuals.
Even this size reduction can only get us so far, however: again with around 10,000 individuals in $I$ and $J$, the problem's size is of order $10^8$, which might still be tractable.
However, larger datasets with hundreds of thousands or even millions of individuals (e.g.\ in the case of exhaustive country data) brings us back into intractable territory.
For these large-scale problems, we still require more tools.

\subsection{Repeated Restricted Optimal Assignment}

In this section we present an algorithm which is able to solve the optimal assignment problem \eqref{eq:optimal_matching_refined} even when $I$ and $J$ are large.
Our procedure not only takes advantage of the structure of the optimal assignment problem to solve it efficiently, it also has an intuitive economic interpretation as a series of two-sided discrete choice problems.
In essence, the algorithm consists of a repeated optimal assignment procedure in which the set of allowed matches, initially restricted, expands iteratively by adding each individual's favorite option.
Our algorithm has links with existing tools in optimization, since it can be seen as a particular case of the Dantzig--Wolfe decomposition (a general algorithm for large-scale linear programming, see \citealt{DantzigWolfe1960}) applied to the assignment problem \eqref{eq:optimal_matching_refined}.

\paragraph{Restricted optimal assignment and choice sets.}
First, we define precisely what we mean by a restricted problem.
For all women $i$, let $Y_i \subset Y$.
We think of $Y_i$ as the subset of men types $y$ that woman $i$ is allowed to match with.
Similarly, let $X_j \subset X$ the subset of women types $x$ that man $j$ is allowed to match with.
We call the sets $Y_i$ and $X_j$ the \emph{choice sets} of woman $i$ and man $j$, respectively.
For any such $(Y_i)$ and $(X_j)$, we define the \emph{restricted optimal assignment problem} associated with the choice sets $(Y_i)$ and $(X_j)$ as
\begin{align}
\label{eq:restricted_problem}
\max_{\substack{
\pi_{iy} \geq 0, \, y \in Y_i \cup \{0\}, \\
\pi_{xj} \geq 0, \, x \in X_j \cup \{0\} }}
& \sum_i \Big[ \pi_{i0} \varepsilon_{i0} + \sum_{y \in Y_i} \pi_{iy} \alpha_{iy} \Big]
+ \sum_j \Big[ \pi_{0j} \eta_{0j} + \sum_{x \in X_j} \pi_{xj} \gamma_{xj} \Big]
\tag{$\mathcal R$}\\
\text{s.t.} ~
& \pi_{i0} + \textstyle\sum_{y \in Y_i} \pi_{iy} = 1 \qquad (\forall i) \notag \\
& \pi_{0j} + \textstyle\sum_{x \in X_j} \pi_{xj} = 1 \qquad (\forall j) \notag \\
& \textstyle\sum_{i : y \in Y_i} \delta_{ix} \pi_{iy} = \textstyle\sum_{j : x \in X_j} \delta_{jy} \pi_{xj} \qquad (\forall xy). \notag
\end{align}
The problem \eqref{eq:restricted_problem} is nothing more than an assignment problem between women $i$ and men $j$, with the twist that agents can only be matched to types in their choice sets.
To see this, observe that \eqref{eq:restricted_problem} is obtained from the assignment problem \eqref{eq:optimal_matching_refined} by adding the constraints that $\pi_{iy} = 0$ whenever $y \notin Y_i$ and $\pi_{xj} = 0$ whenever $x \notin X_j$.

Consider a simple example as illustration.
When choice sets are empty, i.e.\ $Y_i = X_j = \emptyset$ for all $i$ and $j$, problem \eqref{eq:restricted_problem} only features the variables $\pi_{i0}$ and $\pi_{0j}$.
The individual feasibility constraints write as $\pi_{i0} = 1$ and $\pi_{0j} = 1$.
The balance conditions (indexed by $xy$) involve empty sums and are thus trivially verified.
The solution to \eqref{eq:restricted_problem} is thus the matching where every agent remains single.
This is of course consistent with all choice sets being empty, i.e.\ no match being allowed.


We call any solution to the restricted optimal assignment problem \eqref{eq:restricted_problem}, a \emph{restricted optimal matching}.
On the one hand, restricted optimal matchings are typically faster to compute than (unrestricted) optimal matchings, because \eqref{eq:restricted_problem} has fewer variables than \eqref{eq:optimal_matching_refined}.
But on the other hand, it is clear that restricted optimal matchings are \emph{not} optimal in general.
The following result provides a simple criterion to determine when a restricted optimal matching actually coincides with an optimal matching.



\begin{proposition}
\label{prop:restricted_sol_iff_unrestricted_sol}
Let $\pi$ be a restricted optimal matching, and let $u$, $v$, $T$ be the vectors of Lagrange multipliers associated with $\pi$.
Then $\pi$ is an optimal matching if and only if no individual strictly prefers a type outside their choice set, i.e.\
\begin{equation}
\label{eq:dual_feasibility}
u_i \geq \max_{y \in Y \backslash Y_i} ~ \alpha_{iy} - \sum_x \delta_{ix} T_{xy} \quad (\forall i)
\qquad \text{and} \qquad
v_j \geq \max_{x \in X \backslash X_j} ~ \gamma_{xj} + \sum_y \delta_{jy} T_{xy} \quad (\forall j).
\end{equation}
\end{proposition}

The logic behind Proposition \ref{prop:restricted_sol_iff_unrestricted_sol} comes from the standard Walrasian equilibrium interpretation of the assignment problem.
Given the transfers $T_{xy}$ which support the matching, the value $\alpha_{iy} - \sum_i \delta_{ix} T_{xy}$ is the utility woman $i$ would get from matching with a man of type $y$.
The inequality in \eqref{eq:dual_feasibility} thus compares her current utility $u_i$ to her best option among types she is currently \emph{not} allowed to match with, i.e.\ $y \in Y \setminus Y_i$.
The same interpretation applies symmetrically to men.
Hence, condition \eqref{eq:dual_feasibility} imposes that, under the current transfers, everyone is already matched to their favorite type.

\paragraph{Algorithm.}
In fact, Proposition \ref{prop:restricted_sol_iff_unrestricted_sol} does more than certify optimality; it also indicates how to improve the choice sets when optimality fails.
Suppose we start from given choice sets $(Y_i)$ and $(X_j)$, solve the restricted problem \eqref{eq:restricted_problem}, and find that woman $i$ would strictly prefer some type $y \notin Y_i$ under the current transfers.
We should therefore expand her choice set $Y_i$ to include that type $y$.
Symmetrically, if some man $j$ would strictly prefer type $x \notin X_j$, we should add the type $x$ to $X_j$.
Following this logic, we obtain Algorithm 1.

\begin{figure}[h]
\begin{algorithm}[H]
\label{alg:RROA}
\caption{Repeated Restricted Optimal Assignment}
\DontPrintSemicolon
\KwIn{Matching problem $(I, J, X, Y, \delta, \Phi, \varepsilon, \eta)$}
\KwOut{Optimal matching $\pi$}
\emph{Step 0.} Initialize the choice sets as $X_i^0 = \emptyset$ for $i \in I$ and $Y_j^0 = \emptyset$ for $j \in J$. Go to step 1. \;
\emph{Step t.}
\begin{enumerate}[label=(\roman*)]
\item Solve \eqref{eq:restricted_problem} with $Y_i = Y_i^t$ and $X_j = X_j^t$ to obtain a restricted optimal matching $\pi^t$ and its vectors of Lagrange multipliers $u^t, v^t, T^t$.

\item If
\begin{equation*}
u_i^t \geq \max_{y \in Y \setminus Y_i} ~ \alpha_{iy} - \sum_x \delta_{ix} T_{xy}^t \quad (\forall i)
\quad \text{and} \quad
v_j^t \geq \max_{x \in X \setminus X_j} ~ \gamma_{xj} + \sum_y \delta_{jy} T_{xy}^t \quad (\forall j),
\end{equation*}
stop.
Otherwise, go to (iii).

\item For each $i \in I$, if $i$ does not satisfy her above inequality, pick a type $y_i$ in the argmax and define $Y_i^{t+1} = Y_i^t \cup \{y_i\}$; otherwise, define $Y_i^{t+1} = Y_i^t$.
Similarly, for each $j \in J$, if $j$ does not satisfy his above inequality, pick a type $x_j$ in the argmax and define $X_j^{t+1} = X_j^t \cup \{x_j\}$; otherwise, define $X_j^{t+1} = X_j^t$.
Go to step $t+1$.

\end{enumerate}

\Return{$\pi = \pi^t$}
\end{algorithm}
\end{figure}

Notice that in Algorithm \ref{alg:RROA}, the agents' choice sets expand monotonically, hence the algorithm must eventually converge.
At every step, at least one agent whose inequality in \eqref{eq:dual_feasibility} is violated gains a new admissible type in their choice set.
Since there is a finite number of agents and of possible types, this process can only repeat finitely many times; specifically, $|I| |X| + |J| |Y|$ times.
When no choice set expands anymore, all inequalities in \eqref{eq:dual_feasibility} are satisfied, and by Proposition \ref{prop:restricted_sol_iff_unrestricted_sol}, the current restricted optimal matching is globally optimal.
We can thus state the following result.

\begin{proposition}
\label{prop:convergence_RROA}
Algorithm \ref{alg:RROA} terminates in at most $|I| |X| + |J| |Y|$ steps and returns an optimal matching.
\end{proposition}

The monotonicity of the choice sets as a convergence condition is not without recalling Gale--Shapley's deferred acceptance algorithm to find stable matchings \citep{GaleShapley1962}.
In Gale--Shapley however, suitors' choice sets start full and progressively shrink as suitors get turned down by their most preferred match, as opposed to starting empty and expanding in our case.
Of course, another difference is that Gale--Shapley applies to problems with nontransferable utility, whereas our algorithm targets problems with transferable utility.

\subsection{Simulations}
\label{sec:optimal_assignment_simulations}

To conclude this section on the optimal assignment problem, we benchmark the performance of our RROA procedure (Algorithm \ref{alg:RROA}) against a state-of-the-art, general-purpose solver (Gurobi).
We run numerical experiments for different parameter values and population sizes and compare the time needed to reach a solution using these two methods.
Details about the hardware and software specifications we used can be found in Appendix \ref{apx:simulations}.

To compare the two methods, we consider several sizes for the type sets $X$ and $Y$ and for the populations $I$ and $J$.
Then, for given sizes of these sets, we generate 10 random problems and solve them separately using RROA and Gurobi.
For each problem we compute the relative performance of RROA as the ratio of the two solving times.
Finally, we average this relative performance over the 10 trials to obtain a measure of the relative performance for these population and type sets sizes.

By default, Gurobi runs multiple LP solvers concurrently until one of them terminates.
In order to make the comparison between methods easier, in this section we present the results when Gurobi is restricted to use a single method at a time, namely the Dual Simplex and Barrier methods.
These two methods were chosen since they were the two methods consistently used by Gurobi in our experiments.
By doing this, we isolate the RROA's advantage over these families of solvers.
The benchmarking against Gurobi with its default, adaptive method also yielded significant computational improvements; these results are presented in Appendix \ref{apx:simulations}. 

\begin{figure}[p]
    \centering
    \includegraphics[width=0.65\textwidth]{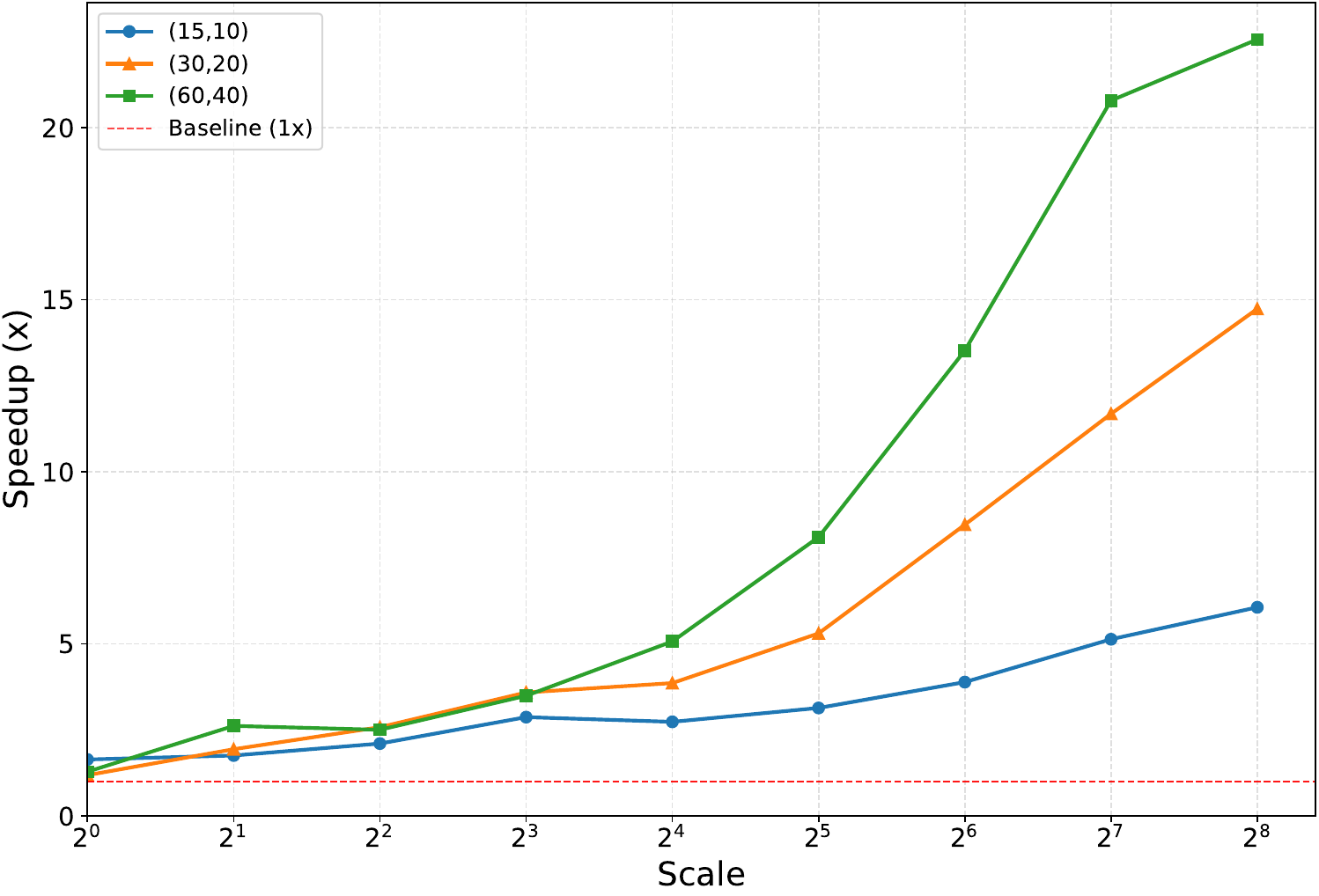}
    \caption{
    Relative performance of the RROA procedure when forcing Gurobi to use the Dual Simplex method.
    Each line corresponds to a fixed value of the size of the type sets $(|X|,|Y|)$.
    The horizontal axis varies the size of the population using an exponential scale as $|I| = 400 \cdot S$ and $|J| = 300 \cdot S$ for $S = 1, 2, 4, \dots, 256$.
    Each point was obtained by generating and solving 10 independent assignment problems for the corresponding value of $(|X|, |Y|, S)$ and by averaging the speedup performance over those 10 problems.
    Assignment problems were generated by independently drawing surpluses $\Phi_{xy}$ from $\mathcal{N}(0, 5^2)$ and idiosyncratic preference shocks from $\mathcal{N}(0, 0.1^2)$.
    }
    \label{fig:speedup-scale}
    \vspace{30pt}
    \includegraphics[width=0.65\textwidth]{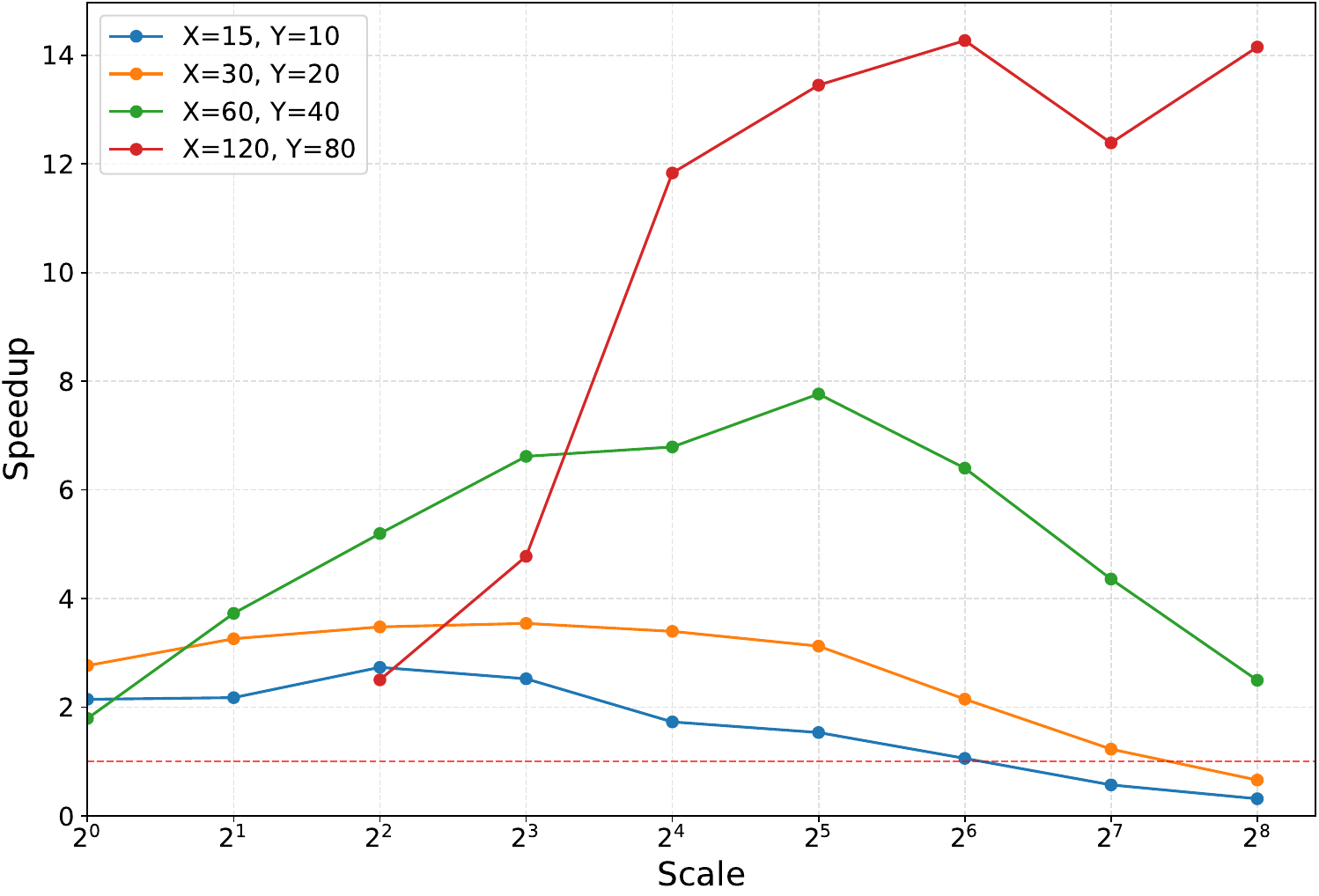}
    \caption{
    Relative performance of the RROA procedure when forcing Gurobi to use the Barrier method.
    The methodology to construct this graph is the same as for Figure \ref{fig:speedup-scale}, except that we forced Gurobi to use a consistent method across population scale factors.
    }
    \label{fig:speedup-scale-forced}
\end{figure}

The results are displayed in Figure \ref{fig:speedup-scale} and \ref{fig:speedup-scale-forced}.
We immediately notice that the computational gains of RROA are highly dependent on the size of the sets $X$, $Y$, $I$, and $J$. Against the Dual Simplex method, RROA's speedups steadily grow as scale increases and eventually achieve up to 25x speedups for larger number of types.
In absolute terms, this corresponded to computing times of around 2 minutes for our RROA method, vs.\ 42 minutes for Gurobi with the Dual Simplex method to solve one problem.
Against the Barrier method, smaller type sets achieve speedups up to 3--5x before declining closer to 1x for larger scales. In contrast, for higher number of types, the advantage becomes substantial, and the algorithm achieves up to 15x speedups.
In absolute terms, this corresponded to computing times of around 90 seconds for our RROA method, vs.\ 23 minutes for Gurobi with the Barrier method to solve one problem.
These results suggest that the RROA method is most efficient on a large scale when there are sufficiently many types. 


We limited the simulations up to a $2^8$ population scale due to hardware computational constraints.
(At the higher scale, solving a single assignment problem with Gurobi lasted around 40min.)
Although we cannot directly perform the numerical experiments for larger population sizes, the steady increase in relative performance for larger number of types strongly suggests that the performance gap would continue to increase as the population size increases further.

Overall, the numerical evidence suggests that our proposed RROA algorithm is highly efficient for solving problems with large population sizes and rich type sets, significantly outperforming state-of-the-art linear programming solvers.

\section{Estimation}
\label{sec:estimation}

In section \ref{sec:optimal_assignment} we saw how to solve the optimal assignment problem for given systematic surpluses $\Phi_{xy}$ and arbitrary idiosyncratic preference shocks $\varepsilon_{iy}$ and $\eta_{xj}$, even as the population size grows large.
In this section we tackle the inverse problem:
Given an observed matching and a distributional assumption on the idiosyncratic shocks, we want to recover the systematic surpluses $\Phi_{xy}$.

We assume the analyst observes a population partitioned into observable types, with $n_x$ women of type $x$ and $m_y$ men of type $y$.
The data consist of an aggregate matching, that is, a matrix $\hat \mu = (\hat \mu_{xy})$ whose entries $\hat \mu_{xy}$ record the number of observed matches between women $x$ and men $y$.
Given these observed matches, the number or singles of each type is therefore $\hat \mu_{x0} = n_x - \sum_y \hat \mu_{xy}$ for women $x$, and $\hat \mu_{0y} = m_y - \sum_x \hat \mu_{xy}$ for men $y$.

In this section, we will also assume that the systematic surplus follows a linear parametrization which depends on a parameter vector $\lambda$.

\begin{assumption}[Linear parametrization of $\Phi$] \label{ass:linear_surplus}
There is a parameter vector $\lambda \in \mathbb R^K$ and a basis of $K$ linearly independent surplus vectors $\phi_k = (\phi_{xyk})$ such that for all $xy$,
\begin{equation}
    \Phi_{xy} = \sum_{k=1}^K \phi_{xyk} \lambda_k,
\end{equation}
or, written in vector form, $\Phi = \phi \lambda$.
\end{assumption}

We assume that the surplus vectors $\phi_k$ are observable.
In practice, these dimensions $k$ could include fixed effects for the types $x$ and $y$, as well as interaction terms representing for instance type proximity.
As long as the basis of surplus vectors is rich enough, it will be able to reconstruct any surplus matrix $\Phi$.



\subsection{Social surplus and entropy of matching}

In order to build our estimator of the systematic surplus, we start by considering a continuous population approximation of the assignment problem \eqref{eq:optimal_matching_refined}.
In this approximation, each individual from the finite population problem is replaced by a unit mass of individuals with the same type.
Given a matrix $\Phi = (\Phi_{xy})$ of systematic surpluses, we then define the social surplus as
\begin{align} \label{eq:social_surplus}
\mathcal W (\Phi) = \max_{\mu_{xy} \geq 0} ~
& \sum_{xy} \mu_{xy} \Phi_{xy} 
+ \mathcal E (\mu) \\
\text{s.t.} ~ & \textstyle\sum_y \mu_{xy} \leq n_x \quad (\forall x) \notag \\
              & \textstyle\sum_x \mu_{xy} \leq m_y \quad (\forall y), \notag
\end{align}
where $\mathcal E (\mu)$ is the \emph{entropy of matching} for the aggregate matching $\mu$,
that is, the maximal contribution of the idiosyncratic shocks which is consistent with $\mu$.
This entropy is additively separable in each type.
This is because, once the aggregate matching $\mu$ is fixed, maximizing the contribution of the idiosyncratic shocks boils down to assigning individuals within each observable type to singlehood or to partner types.



For example, consider women of type $x$.
The aggregate matching vector $\mu$ tells us exactly how much mass of women $x$ must match with men $y$, and how much must remain single.
The only freedom left is therefore to assign each woman $i$ of type $x$ across these options in a way that maximizes the total contribution of their idiosyncratic shocks, while respecting the aggregate masses $\mu_{xy}$.
With this logic, the expression for the entropy of matching for women $x$ is
\begin{align} \label{eq:entropy_of_matching_x}
\mathcal E_x (\mu) = \sup_{\pi_{i0}, \pi_{iy} \geq 0} ~ & \int_i \Big( \pi_{i0} \varepsilon_{i0} + \sum_y \pi_{iy} \varepsilon_{iy} \Big) \, \text d\mathbf P_x (\varepsilon_i) \\
\text{s.t.} ~ & \textstyle \pi_{i0} + \sum_y \pi_{iy} = 1 \quad (\forall i) \notag \\
              & \textstyle \int_i \pi_{iy} \, \text d \mathbf P_x (\varepsilon_i) = \mu_{xy} \quad (\forall y). \notag
\end{align}
The entropy of matching $\mathcal E_y (\mu)$ for men $y$ has an analogous expression.
The total entropy of matching is then obtained by adding the entropy of matching for all types,
\[
\mathcal E (\mu) = \sum_x \mathcal E_x (\mu) + \sum_y \mathcal E_y (\mu).
\]

In a few cases, the entropy of matching has an explicit analytical expression.
In particular, when $\mathbf P_x$ and $\mathbf Q_y$ are the distributions of i.i.d.\ Gumbel preference shocks, $\mathcal E (\mu)$ is the usual entropy.%
\footnote{That is to say, $\mathcal E (\mu) = \sum_{xy} \mu_{xy} \ln \mu_{xy}$ up to a constant. See \citet{GalichonSalanie2022} for details.}
However, for general distributions there is no simple expression for the entropy of matching.
In this case, we can instead approximate it using a sample equivalent of \eqref{eq:entropy_of_matching_x}.
Specifically, we draw a sample $(\varepsilon_i)$ of size $n_x$ from $\mathbf P_x$ and compute the sample equivalent of \eqref{eq:entropy_of_matching_x} as
\begin{align*}
\widehat{\mathcal E}_x (\mu)
= \max_{\pi_{i0}, \pi_{iy} \geq 0} ~ &
\sum_i \Big[ \pi_{i0} \varepsilon_{i0} + \sum_y \pi_{iy} \varepsilon_{iy} \Big] \\
\text{s.t.} ~ & \textstyle \pi_{i0} + \sum_y \pi_{iy} = 1 \quad (\forall i) \\
              & \textstyle \sum_i \pi_{iy} = \mu_{xy} \quad (\forall y).
\end{align*}
We similarly obtain $\widehat{\mathcal E}_y (\mu)$ by drawing a sample $(\eta_j)$ of size $m_y$ from $\mathbf Q_y$.
We then define the 
\emph{simulated entropy of matching} as
$\widehat{\mathcal E} (\mu) = \sum_x \widehat{\mathcal E}_x (\mu) + \sum_y \widehat{\mathcal E}_y (\mu)$, which we can rewrite as a single maximization program:
\begin{align} \label{eq:simulated_entropy_of_matching}
\widehat{\mathcal E} (\mu)
= \max_{\pi_{i0}, \pi_{iy}, \pi_{0j}, \pi_{xj} \geq 0} ~ &
\sum_i \Big[ \pi_{i0} \varepsilon_{i0} + \sum_y \pi_{iy} \varepsilon_{iy} \Big]
+ \sum_j \Big[ \pi_{0j} \eta_{0j} + \sum_x \pi_{xj} \eta_{xj} \Big] \\
\text{s.t.} ~ & \textstyle \pi_{i0} + \sum_y \pi_{iy} = 1 \quad (\forall i) \notag \\
              & \textstyle \pi_{0j} + \sum_x \pi_{xj} = 1 \quad (\forall j) \notag \\
              & \textstyle \sum_i \delta_{ix} \pi_{iy} = \mu_{xy} \quad (\forall xy) \notag \\
              & \textstyle \sum_j \delta_{jy} \pi_{xj} = \mu_{xy} \quad (\forall xy). \notag
\end{align}
In turn, we define the simulated social surplus $\widehat{\mathcal W} (\Phi)$ by replacing the entropy $\mathcal E (\mu)$ by its simulated counterpart $\widehat{\mathcal E} (\mu)$ in the definition \eqref{eq:social_surplus} of the social surplus $\mathcal W (\Phi)$.
As expected, the simulated social surplus simply corresponds to the value of the optimal assignment problem associated with the simulated population.

\begin{proposition} \label{prop:simulated_social_surplus_is_optimal_assignment}
For given draws $(\varepsilon_i)$ and $(\eta_j)$ of the idiosyncratic preference shocks, the simulated social surplus $\widehat{\mathcal W} (\Phi)$ is equal to the value of the optimal assignment problem \eqref{eq:optimal_matching_refined}.
\end{proposition}

Since we are interested in general distributions $\mathbf P_x$ and $\mathbf Q_y$, and therefore cannot rely on the entropy having an analytical expression, the simulated social surplus will play an important role in our estimation procedure.
In this respect, Proposition \ref{prop:simulated_social_surplus_is_optimal_assignment} already hints at the fact that the computation method developed for solving the optimal assignment problem in section \ref{sec:optimal_assignment} will be useful for estimation as well.

\subsection{Simulated moment-matching estimator}

In this section we consider that Assumption \ref{ass:linear_surplus} holds, so that the surplus is parametrized as $\Phi = \phi \lambda$, where $\phi = (\phi_{xyk})$ is observed by the analyst and $\lambda = (\lambda_k)$ is a vector of parameters to be estimated.
From the social surplus \eqref{eq:social_surplus}, we derive a method of moments estimator for $\lambda$ which will serve as the basis for our estimation procedure.
Denote $\mu^\lambda$ the solution to \eqref{eq:social_surplus} when $\Phi = \phi \lambda$.
The envelope theorem applied to $\mathcal W(\phi \lambda)$ yields
\[
\phi^\top \nabla \mathcal W(\phi \lambda) = \phi^\top \mu^\lambda.
\]
It is therefore natural to consider the estimator tied to the moment conditions $\phi^\top \mu^\lambda = \phi^\top \hat \mu$.
Furthermore, observe that these moment conditions are simply the first-order conditions of the convex optimization problem:
\begin{equation} \label{eq:method_of_moments}
    \max_\lambda ~ (\phi \lambda)^\top \hat \mu - \mathcal W (\phi \lambda).
\end{equation}
The solution to problem \eqref{eq:method_of_moments} is thus the moment-matching estimator for $\lambda$.

As the solution to a convex optimization problem, the moment-matching estimator should in theory be straightforward to compute.
Recall, however, that the social surplus function $\mathcal W (\phi \lambda)$ is itself obtained as the value of an optimization problem.
Moreover, it involves the entropy of matching $\mathcal E (\mu)$ which, as we discussed above, typically does not have an analytical expression.
For this reason, we will focus on a simulated moment-matching estimator, which is obtained by replacing these quantities with their simulated counterparts introduced in the previous section.

\begin{definition}
The simulated moment-matching (SMM) estimator $\hat \lambda$ is the solution to
\begin{equation} \label{eq:simulated_method_of_moments}
    \max_\lambda ~ (\phi \lambda)^\top \hat \mu - \widehat{\mathcal W} (\phi \lambda).
\end{equation}
It satisfies the moment-matching conditions $\phi^\top \mu^\lambda = \phi^\top \hat \mu$, where $\mu^\lambda$ is obtained by aggregating the solution to the optimal assignment problem \eqref{eq:optimal_matching_refined} for $\Phi = \phi \lambda$ and for the simulated $(\varepsilon_i)$ and $(\eta_j)$. 
\end{definition}

Observe that the definition of the SMM estimator in fact relies of three nested optimization problems: the outer problem in \eqref{eq:simulated_method_of_moments}, an intermediate one in the definition of $\widehat{\mathcal W} (\phi \lambda)$, and an inner one in the expression \eqref{eq:simulated_entropy_of_matching} of $\widehat{\mathcal E} (\mu)$.
By collapsing these three optimization problems into a single one, we obtain the following result.


\begin{proposition} \label{prop:SMM_program}
The simulated moment-matching estimator $\hat \lambda = (\hat \lambda_k)$ is obtained as the vector of Lagrange multipliers of the constraints indexed by $k$ in the following linear program:
\begin{align} \label{eq:SMM_estimation}
\max_{\pi_{iy}, \pi_{i0}, \pi_{xj}, \pi_{0j} \geq 0} ~&
\sum_i \Big( \pi_{i0} \varepsilon_{i0} + \sum_y \pi_{iy} \varepsilon_{iy} \Big) + \sum_j \Big( \pi_{0j} \eta_{0j} + \sum_x \pi_{xj} \eta_{xj} \Big) \\
\text{s.t.} ~ & \pi_{i0} + \textstyle \sum_y \pi_{iy} = 1 \qquad (\forall i) \notag \\
              & \pi_{0j} + \textstyle \sum_x \pi_{xj} = 1 \qquad (\forall j) \notag \\
              & \textstyle \sum_i \delta_{ix} \pi_{iy} = \sum_j \delta_{jy} \pi_{xj} \qquad (\forall xy) \notag \\
              & \textstyle \sum_{xy} \frac{1}{2} \big(\sum_i \delta_{ix} \pi_{iy} + \sum_j \delta_{jy} \pi_{xj} \big) \phi_{xyk} = \sum_{xy} \hat \mu_{xy} \phi_{xyk} \qquad (\forall k). \notag
\end{align}
\end{proposition}

Proposition \ref{prop:SMM_program} states that the SMM estimator is obtained by solving a linear program which closely resembles an optimal assignment problem.
Indeed, the program \eqref{eq:SMM_estimation} differs from \eqref{eq:optimal_matching_refined} in only two ways.
First, its objective only includes the individual heterogeneity components of the utility from matching, and not the systematic utilities obtained from splitting the systematic surplus.
Second, it includes one more set of constraints (those indexed by $k$), which exactly correspond to the $k$ moment-matching conditions of our estimator.

\subsection{Computation}

We now use the result of Proposition \ref{prop:SMM_program} as the basis for a method to compute the SMM estimator $\hat \lambda$.
Since the linear program \eqref{eq:SMM_estimation} closely resembles the optimal assignment problem, we can reasonably expect that a method similar to Algorithm \ref{alg:RROA} would work as well.
This is in fact the case: after doing the legwork of building the RROA algorithm for the optimal assignment problem, it now suffices to adjust that method while accounting for the modified objective and new set of constraints.

As in section \ref{sec:optimal_assignment}, we consider choice sets $Y_i$ and $X_j$ for all individuals.
We then define the restricted version of the estimation problem \eqref{eq:SMM_estimation} associated with such choice sets:
\begin{align} \label{eq:SMM_estimation_restricted}
\max_{\substack{
\pi_{iy} \geq 0, \, y \in Y_i \cup \{0\}, \\
\pi_{xj} \geq 0, \, x \in X_j \cup \{0\} }} ~&
\sum_i \Big( \pi_{i0} \varepsilon_{i0} + \sum_{y \in Y_i} \pi_{iy} \varepsilon_{iy} \Big) + \sum_j \Big( \pi_{0j} \eta_{0j} + \sum_{x \in X_j} \pi_{xj} \eta_{xj} \Big) \\
\text{s.t.} ~ & \pi_{i0} + \textstyle \sum_y \pi_{iy} = 1 \qquad (\forall i) \notag \\
              & \pi_{0j} + \textstyle \sum_x \pi_{xj} = 1 \qquad (\forall j) \notag \\
              & \textstyle \sum_i \delta_{ix} \pi_{iy} = \sum_j \delta_{jy} \pi_{xj} \qquad (\forall xy) \notag \\
              & \textstyle \sum_{xy} \frac{1}{2} \big(\sum_i \delta_{ix} \pi_{iy} + \sum_j \delta_{jy} \pi_{xj} \big) \phi_{xyk} = \sum_{xy} \hat \mu_{xy} \phi_{xyk} \qquad (\forall k). \notag
\end{align}
By a reasoning similar to that developed in section \ref{sec:optimal_assignment}, we obtain Algorithm \ref{alg:RROA_estimation}.

\vspace{10pt}

\begin{algorithm}[H]
\label{alg:RROA_estimation}
\caption{Repeated Restricted Optimal Assignment for the SMM estimator}
\DontPrintSemicolon
\KwIn{Observed matching $\hat \mu$ and characteristics $\phi$}
\KwOut{SMM estimator $\hat \lambda$}
\emph{Step 0.}
For all $x$, simulate a sample $(\varepsilon_i)$ of size $n_x$ from $\mathbf P_x$;
and for all $y$, simulate a sample $(\eta_j)$ of size $m_y$ from $\mathbf Q_y$.
Initialize the choice sets as $X_i^0 = \emptyset$ for $i \in I$ and $Y_j^0 = \emptyset$ for $j \in J$. Go to step 1. \;
\emph{Step t.}
\begin{enumerate}[label=(\roman*)]
\item Solve \eqref{eq:SMM_estimation_restricted} with $Y_i = Y_i^t$ and $X_j = X_j^t$ to obtain a restricted optimal matching $\pi^t$ and its vectors of Lagrange multipliers $u^t, v^t, T^t, \lambda^t$.

\item If \vspace{-20pt}
\begin{align*}
&u_i^t \geq \max_{y \in Y \setminus Y_i} ~ \sum_x \delta_{ix} \Big(\frac{1}{2} \sum_k \phi_{xyk} \lambda_k^t - T_{xy}^t \Big) + \varepsilon_{iy} \quad (\forall i) \\
\text{and} \quad
&
v_j^t \geq \max_{x \in X \setminus X_j} ~ \sum_y \delta_{jy} \Big(\frac{1}{2} \sum_k \phi_{xyk} \lambda_k^t + T_{xy}^t\Big) + \eta_{xj} \quad (\forall j),
\end{align*}
stop.
Otherwise, go to (iii).

\item For each $i \in I$, if $i$ does not satisfy her above inequality, pick a type $y_i$ in the argmax and define $Y_i^{t+1} = Y_i^t \cup \{y_i\}$; otherwise, define $Y_i^{t+1} = Y_i^t$.
Similarly, for each $j \in J$, if $j$ does not satisfy his above inequality, pick a type $x_j$ in the argmax and define $X_j^{t+1} = X_j^t \cup \{x_j\}$; otherwise, define $X_j^{t+1} = X_j^t$.
Go to step $t+1$.

\end{enumerate}

\Return{$\hat \lambda = \lambda^t$}
\end{algorithm}

\vspace{10pt}

Similar to Algorithm \ref{alg:RROA}, this version terminates in at most $|I| |Y| + |J| |X|$ steps.





\subsection{Simulations}

We now present the accuracy of the estimation using the RROA algorithm.
Since the structure of the problem is quite similar to that of solving the optimal assignment problem, the computational results are analogous from those of section \ref{sec:optimal_assignment_simulations}.
We thus focus instead on the consistency of our estimator as the population size grows. 

As in section \ref{sec:optimal_assignment_simulations}, we run numerical experiments for different sizes of the population $I$ and $J$.
We fixed the size of the basis of surplus vectors to $K=5$.
For each simulation, a true parameter vector $\lambda$ was simulated, as well as a basis of surplus vectors $\phi$, and finally idiosyncratic shocks drawn i.i.d.\ from $\mathcal N (0, 10^{-2})$.
We then solved the optimal assignment problem associated with the surplus matrix $\Phi = \phi \lambda$ over these simulated individuals using Algorithm \ref{alg:RROA}, yielding an optimal matching.
Aggregating it, we obtained a matrix $\hat \mu$ which can serve as the matrix of observed matches for the estimation.

We then estimate $\lambda$ using this data by redrawing a new set of shocks.
In one case, we draw these shocks from the accurate data generating distribution $\mathcal N (0, 10^{-2})$.
In another case, we draw them from a misspecified distribution, namely a Gumbel distribution with same mean and variance.
We then compute our estimator of $\lambda$ by solving \eqref{eq:SMM_estimation} in those two cases.
We finally compute the Normalized Root Mean Square Error (NRMSE) of our two estimators, which is a measure of the distance from the estimator to the true value of $\lambda$.
We repeat this process 100 times for each value of $I,J$, and compute the average NRMSE over these 100 trials.

\begin{figure}[ht]
    \centering
    \includegraphics[width=0.85\textwidth]{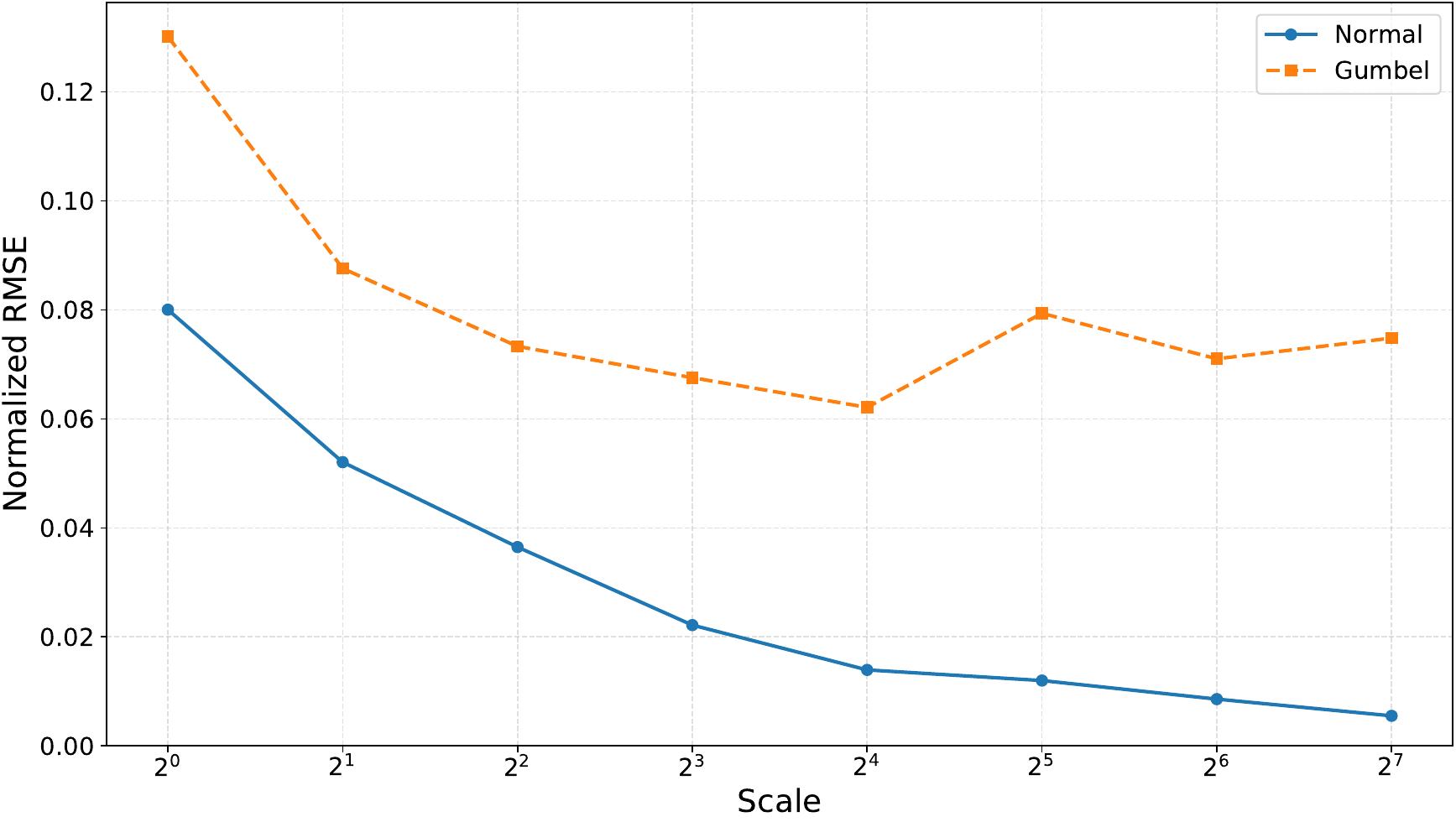}
    \caption{
    Empirical consistency of the estimator.
    Each point represents the Normalized Root Mean Square Error of the estimator of $\lambda$ compared to its true value, averaged over 100 random trials.
    The horizontal axis varies the total size of the population as $|I| = 400 \cdot S$ and $|J| = 300 \cdot S$ for $S = 1, 2, 4, ..., 256$.
    The size of the type sets is fixed to $|X| = 15$, $|Y| = 10$.
    }
    \label{fig:estimation-error}
\end{figure}

The results of these simulations are displayed in Figure \ref{fig:estimation-error}.
We observe that the estimation accuracy improves steadily as population size grows, suggesting that our estimator is consistent.
When the model is well-specified, that is, when the errors used in the estimation are drawn from the true normal distribution, the NRMSE eventually drops below 1\% for the highest population size we investigated.
When, instead, the model is misspecified and errors are drawn from a Gumbel distribution, the NRMSE stabilizes at around 7\%, suggesting a persistent bias due to the misspecification.

\section{Conclusion}
\label{sec:conclusion}

We develop a tractable framework for the empirical analysis of matching markets with transferable utility that dispenses with the i.i.d.\ Gumbel assumption while preserving the separability structure of \citet{ChooSiow2006}.
On the computational side, our Repeated Restricted Optimal Assignment (RROA) algorithm solves large assignment problems orders of magnitude faster than off-the-shelf LP solvers. On the econometric side, we show how the same structure yields a simulated moment-matching estimator that remains feasible under general error distributions, and we document (i) consistency under correct specification and (ii) systematic bias when a logit error is misspecified for probit.

Future work will first investigate probit specifications in more details,with rich cross-attribute covariance (including factor structures), and quantify how correlation shapes substitution patterns and welfare relative to logit.
Second, it will replicate canonical TU matching results (education, income, ethnicity) under probit and nested-logit errors to assess the robustness of estimated sorting and counterfactuals.
Third, it will develop inference (standard errors, over-identification tests) for the simulated moment-matching estimator.

\newpage

\bibliographystyle{agsm}
\bibliography{references}

@article{Shapley-Shubik1971,
  title={The assignment game {I}: the core},
  author={L. S. Shapley and M. Shubik},
  journal={International Journal of Game Theory},
  volume={1},
  number={},
  pages={111--130},
  year={1971}
}

@article{ChooSiow2006,
  title={Who Marries Whom and Why},
  author={Choo, E. and Siow, A.},
  journal={Journal of Political Economy},
  volume={114},
  number={},
  pages={175--201},
  year={2006}
}

@article{GaleShapley1962,
  title={College admissions and the stability of marriage},
  author={Gale, D. and Shapley, L. S.},
  journal={The American Mathematical Monthly},
  volume={69},
  number={1},
  pages={9--15},
  year={1962}
}

@article{GalichonSalanie2022,
  title={Cupid’s Invisible Hand: Social Surplus and Identification in Matching Models},
  author={Alfred Galichon and Bernard Salanié},
  journal={The Review of Economic Studies},
  volume={89},
  number={5},
  pages={2600--2629},
  year={2022}
}

@article{ChiapporiSalanieWeiss2017,
  title={Partner Choice, Investment in Children, and the Marital College Premium},
  author={Pierre-André Chiappori and Bernard Salanié and Yoram Weiss},
  journal={The American Economic Review},
  volume={107},
  number={8},
  pages={2109--67},
  year={2017}
}

@Article{DantzigWolfe1960,
  author   = {George B. Dantzig and Philip Wolfe},
  journal  = {Operations Research},
  title    = {Decomposition Principle for Linear Programs},
  year     = {1960},
  pages    = {101--111},
  volume   = {8}
}

@Article{GousseJacquemetRobin2017,
  author   = {Marion Goussé and Nicolas Jacquemet and Jean-Marc Robin},
  journal  = {Econometrica},
  title    = {Marriage, labor supply, and home
production},
  year     = {2017},
  pages    = {1873--1919},
  volume   = {85},
  number   = {6}
}

@Article{Ciscato2025,
  author   = {Edoardo Ciscato},
  journal  = {The Review of Economic Studies},
  title    = {Assessing Racial and Educational Segmentation in Large Marriage Markets},
  year     = {2025},
  pages    = {3788--3839},
  volume   = {92},
  number   = {6}
}

@Article{GualdaniSinha2023,
  author   = {Cristina Gualdani and Shruti Sinha},
  journal  = {Journal of Political Economy},
  title    = {Partial Identification in Matching Models for the Marriage Market},
  year     = {2023},
  pages    = {1109-1171},
  volume   = {131},
  number   = {5}
}

@Article{McFadden1989,
  author   = {McFadden, Daniel},
  journal  = {Econometrica},
  title    = {A method of simulated moments for estimation of discrete response models without numerical integration},
  year     = {1989},
  pages    = {995-1026},
}

@Book{Train2009,
  author   = {Train, Kenneth E.},
  journal  = {Econometrica},
  title    = {Discrete choice methods with simulation},
  year     = {2009},
  publisher = {Cambridge university press},
}

@incollection{McFadden1981,
    author = {McFadden, Daniel},
    title  = {Econometric models of probabilistic choice},
    booktitle = {Structural analysis of discrete data with econometric applications},
    number = {198272},
    year = {1981}
}

@Unpublished{ChiapporiNguyenSalanie2019,
  author   = {Pierre-André Chiappori and D. L. Nguyen and Bernard Salanié},
  note     = {Columbia University Mimeo},
  title    = {Matching with Random Components: Simulations},
  year     = {2019},
}

@Unpublished{BisinTura2019,
  author = {Alberto Bisin and Giulia Tura},
  note   = {NBER Working paper 26303},
  title  = {Marriage, Fertility, and Cultural Integration in {Italy}},
  year   = {2019},
}

\newpage
\appendix

\section*{\huge Appendix}

\section{Proofs}
\label{apx:proofs}

\subsection{Proof of Proposition \ref{prop:optimal_assignment_reformulation}}

Introduce $\pi_{i0}$ and $\pi_{0j}$ the singlehood indicators for $i$ and $j$ respectively (here, the slack variables of the linear program), use the separability assumption \ref{ass:separability}, and add the constant $\sum_i \varepsilon_{i0} + \sum_j \eta_{0j}$ to the objective of \eqref{eq:optimal_matching_naive} to rewrite it as
\begin{align*}
\max_{\tilde \pi_{ij}, \pi_{i0}, \pi_{0j \geq 0}} ~ &
\sum_i \sum_j \tilde \pi_{ij} \Big(\frac{\Phi_{x_i y_j}}{2} + \varepsilon_{i y_j}\Big)
+ \sum_j \sum_i \tilde \pi_{ij} \Big(\frac{\Phi_{x_i y_j}}{2} + \eta_{x_i j}\Big)
+ \sum_i \pi_{i0} \varepsilon_{i0} + \sum_j \pi_{0j} \varepsilon_{i0} \\
\text{s.t.} ~ & \pi_{i0} + \textstyle\sum_j \tilde \pi_{ij} = 1 \quad (\forall i), \qquad
                \pi_{0j} + \textstyle\sum_i \tilde \pi_{ij} = 1 \quad (\forall j).
\end{align*}
We then obtain \eqref{eq:optimal_matching_refined} with the change of variables $\pi_{iy} = \sum_{j: y_j = y} \tilde \pi_{ij}$ and $\pi_{xj} = \sum_{i: x_i = x} \tilde \pi_{ij}$,
which implies a new balance condition for matches of type $xy$,
$\sum_{i: x_i = x} \pi_{iy} = \sum_{ij: x_i = x, y_j = y} \tilde \pi_{ij} = \sum_{j: y_j = y} \pi_{xj}$.
\qed

\subsection{Proof of Proposition \ref{prop:restricted_sol_iff_unrestricted_sol}}

Let $\pi$ be a solution to the restricted assignment problem \eqref{eq:restricted_problem} with $u$, $v$ and $T$ its vectors of Lagrange multipliers.
The unrestricted assignment problem \eqref{eq:optimal_matching_refined} has dual
\begin{align} \label{eq:optimal_matching_refined_dual}
\min_{u_i, v_j, T_{xy}} ~ & \sum_i u_i + \sum_j v_j \\
\text{s.t.} ~ & u_i \geq \varepsilon_{i0} \notag               &&(\forall i \in I) \\
              & v_j \geq \eta_{0j} \notag                      &&(\forall j \in J) \\
              & u_i \geq \alpha_{iy} - \textstyle \sum_x \delta_{ix} T_{xy} &&(\forall i \in I, y \in Y) \notag \\
              & v_j \geq \gamma_{xj} + \textstyle \sum_y \delta_{jy} T_{xy} &&(\forall j \in J, x \in X), \notag
\end{align}
so clearly, for $(\pi, u, v, T)$ to be a primal--dual solution to \eqref{eq:optimal_matching_refined} and \eqref{eq:optimal_matching_refined_dual}, it must satisfy the inequalities in \eqref{eq:dual_feasibility}.

Conversely, assume that $u$, $v$, and $T$ satisfy the inequalities in \eqref{eq:dual_feasibility}.
First, note that $\pi$ being solution to the restricted problem \eqref{eq:restricted_problem} implies that it is feasible for the unrestricted problem \eqref{eq:optimal_matching_refined}.
Second, the restricted assignment problem \eqref{eq:restricted_problem} has dual
\begin{align} \label{eq:restricted_problem_dual}
\min_{u_i, v_j, T_{xy}} ~ & \sum_i u_i + \sum_j v_j \\
\text{s.t.} ~ & u_i \geq \varepsilon_{i0} \notag               &&(\forall i \in I) \\
              & v_j \geq \eta_{0j} \notag                      &&(\forall j \in J) \\
              & u_i \geq \alpha_{iy} - \textstyle \sum_x \delta_{ix} T_{xy} &&(\forall i \in I, y \in Y_i) \notag \\
              & v_j \geq \gamma_{xj} + \textstyle \sum_y \delta_{jy} T_{xy} &&(\forall j \in J, x \in X_j). \notag
\end{align}
and since $(u, v, T)$ is solution to \eqref{eq:restricted_problem_dual} and also satisfies \eqref{eq:dual_feasibility}, $(u, v, T)$ is actually feasible for the unrestricted dual \eqref{eq:optimal_matching_refined_dual} as well.
Finally, because $(\pi, u, v, T)$ is solution to the restricted primal--dual problem \eqref{eq:restricted_problem}--\eqref{eq:restricted_problem_dual}, we must have
\begin{align*}
\sum_i u_i + \sum_j v_j
&= \sum_i \Big[ \pi_{i0} \varepsilon_{i0} + \sum_{y \in Y_i} \pi_{iy} \alpha_{iy} \Big]
+ \sum_j \Big[ \pi_{0j} \eta_{0j} + \sum_{x \in X_j} \pi_{xj} \gamma_{xj} \Big] \\
&= \sum_i \Big[ \pi_{i0} \varepsilon_{i0} + \sum_{y \in Y} \pi_{iy} \alpha_{iy} \Big]
+ \sum_j \Big[ \pi_{0j} \eta_{0j} + \sum_{x \in X} \pi_{xj} \gamma_{xj} \Big],
\end{align*}
where the second inequality is due to the fact that $\pi_{iy} = 0$ for any pair $iy$ such that $y \notin Y_i$, and similarly $\pi_{xj} = 0$ for any pair $xj$ such that $x \notin X_j$.
Hence $(\pi, u, v, T)$ is a primal--dual solution for the unrestricted problem.
\qed

\subsection{Proof of Proposition \ref{prop:simulated_social_surplus_is_optimal_assignment}}

By definition, the simulated social surplus is 
\begin{align*}
\widehat {\mathcal W} (\Phi) = \max_{\mu_{xy} \geq 0} ~
& \sum_{xy} \mu_{xy} \Phi_{xy} 
+ \widehat{\mathcal E} (\mu) \\
\text{s.t.} ~ & \textstyle\sum_y \mu_{xy} \leq n_x \quad (\forall x) \\
              & \textstyle\sum_x \mu_{xy} \leq m_y \quad (\forall y).
\end{align*}
Replacing the simulated entropy of matching $\widehat{\mathcal E} (\mu)$ by its own definition  \eqref{eq:simulated_entropy_of_matching} in this expression, we can merge the two maximization programs to obtain
\begin{align*}
\widehat {\mathcal W} (\Phi) = \max_{\mu_{xy}, \pi_{i0}, \pi_{iy}, \pi_{0j}, \pi_{xj} \geq 0} ~
& \sum_{xy} \mu_{xy} \Phi_{xy} +\sum_i \Big[ \pi_{i0} \varepsilon_{i0} + \sum_y \pi_{iy} \varepsilon_{iy} \Big]
+ \sum_j \Big[ \pi_{0j} \eta_{0j} + \sum_x \pi_{xj} \eta_{xj} \Big] \\
\text{s.t.} ~ & \textstyle\sum_y \mu_{xy} \leq n_x \quad (\forall x) \\
              & \textstyle\sum_x \mu_{xy} \leq m_y \quad (\forall y) \\
              & \textstyle \pi_{i0} + \sum_y \pi_{iy} = 1 \quad (\forall i) \\
              & \textstyle \pi_{0j} + \sum_x \pi_{xj} = 1 \quad (\forall j) \\
              & \textstyle \sum_i \delta_{ix} \pi_{iy} = \mu_{xy} \quad (\forall xy) \\
              & \textstyle \sum_j \delta_{jy} \pi_{xj} = \mu_{xy} \quad (\forall xy).
\end{align*}
Next, we remark that $\mu_{xy}$ is a redundant variable since we can substitute it out using the consistency constraints.
We substitute $\mu_{xy}$ in the objective by $\frac{1}{2}\sum_i \delta_{ix} \pi_{iy}+\frac{1}{2}\sum_j \delta_{jy} \pi_{xj}$.
Furthermore, we use the consistency constraints to substitute $\mu_{xy}$ in the population margin constraints.
We get:
\begin{align*}
\widehat {\mathcal W} (\Phi) = \max_{\pi_{i0}, \pi_{iy}, \pi_{0j}, \pi_{xj} \geq 0} ~
& \frac{1}{2}\sum_{xy} \Big[\sum_i \delta_{ix} \pi_{iy} \Big] \Phi_{xy}+\frac{1}{2}\sum_{xy} \Big[\sum_j \delta_{jy} \pi_{xj} \Big] \Phi_{xy} + \notag \\ & + \sum_i \Big[ \pi_{i0} \varepsilon_{i0} + \sum_y \pi_{iy} \varepsilon_{iy} \Big]
+ \sum_j \Big[ \pi_{0j} \eta_{0j} + \sum_x \pi_{xj} \eta_{xj} \Big] \notag \\
\text{s.t.} ~ & \textstyle\sum_y \sum_i \delta_{ix} \pi_{iy} \leq n_x \quad (\forall x) \\
              & \textstyle\sum_x \sum_j \delta_{jy} \pi_{xj} \leq m_y \quad (\forall y) \\
            & \textstyle \pi_{i0} + \sum_y \pi_{iy} = 1 \quad (\forall i) \\
              & \textstyle \pi_{0j} + \sum_x \pi_{xj} = 1 \quad (\forall j) \\
              & \textstyle \sum_i \delta_{ix} \pi_{iy} = \sum_j \delta_{jy} \pi_{xj} \quad (\forall xy). 
\end{align*}
We see that the first two population constraints are now redundant as they are implied by the individual feasibility constraints (since $\sum_i \delta_{ix} = n_x$ and $\sum_j \delta_{jy} = m_y$).
Rearranging the objective and using the definition \eqref{eq:values_match} of $\alpha_{iy}$ and $\gamma_{xj}$, we finally obtain
\begin{align*}
\widehat {\mathcal W} (\Phi) = \max_{\pi_{iy}, \pi_{i0}, \pi_{xj}, \pi_{0j} \geq 0} ~ & \sum_i \bigg[ \pi_{i0} \varepsilon_{i0} + \sum_{y} \pi_{iy} \alpha_{iy} \bigg]
+ \sum_j \bigg[ \pi_{0j} \eta_{0j} + \sum_{x} \pi_{xj} \gamma_{xj} \bigg] \\
\text{s.t.} ~ & \textstyle \pi_{i0} + \sum_y \pi_{iy} = 1 \quad (\forall i) \\
              & \textstyle \pi_{0j} + \sum_x \pi_{xj} = 1 \quad (\forall j) \\
              & \textstyle \sum_i \delta_{ix} \pi_{iy} = \sum_j \delta_{jy} \pi_{xj} \quad (\forall xy) 
\end{align*}
which is our result.
\qed

\subsection{Proof of Proposition \ref{prop:SMM_program}}

Using the parametrization $\Phi = \phi \lambda$ from Assumption \ref{ass:linear_surplus} and the result from Proposition \ref{prop:simulated_social_surplus_is_optimal_assignment}, the simulated social surplus is
\begin{align*}
\widehat {\mathcal W} (\phi \lambda)
= \max_{\pi_{iy}, \pi_{i0}, \pi_{xj}, \pi_{0j} \geq 0} ~
& \sum_i \bigg[ \pi_{i0} \varepsilon_{i0} + \sum_{y} \pi_{iy} \Big(\varepsilon_{iy} + \frac{1}{2} \sum_{xk} \delta_{ix} \phi_{xyk} \lambda_k \Big) \bigg] \\
& + \sum_j \bigg[ \pi_{0j} \eta_{0j} + \sum_{x} \pi_{xj} \Big( \eta_{xj} + \frac{1}{2} \sum_{yk} \delta_{jy} \phi_{xyk} \lambda_k \Big) \bigg] \\
\text{s.t.} ~ & \textstyle \pi_{i0} + \sum_y \pi_{iy} = 1 \quad (\forall i) \\
              & \textstyle \pi_{0j} + \sum_x \pi_{xj} = 1 \quad (\forall j) \\
              & \textstyle \sum_i \delta_{ix} \pi_{iy} = \sum_j \delta_{jy} \pi_{xj} \quad (\forall xy). 
\end{align*}
Substituting this expression inside the program \eqref{eq:simulated_method_of_moments} which yields the SMM estimator and rearranging terms, we can rewrite that program as
\begin{align*}
\max_\lambda \min_{\pi_{iy}, \pi_{i0}, \pi_{xj}, \pi_{0j} \geq 0} ~
& -\sum_i \bigg[ \pi_{i0} \varepsilon_{i0} + \sum_{y} \pi_{iy} \varepsilon_{iy} \bigg]
- \sum_j \bigg[ \pi_{0j} \eta_{0j} + \sum_{x} \pi_{xj} \eta_{xj} \bigg] \\
& + \sum_k \lambda_k \bigg[ \sum_{xy} \hat \mu_{xy} \phi_{xyk} - \sum_{xy} \frac{1}{2} \Big( \sum_i \delta_{ix} \pi_{iy} + \sum_j \delta_{jy} \pi_{xj} \Big) \phi_{xyk} \bigg] \\
\text{s.t.} ~ & \textstyle \pi_{i0} + \sum_y \pi_{iy} = 1 \quad (\forall i) \\
              & \textstyle \pi_{0j} + \sum_x \pi_{xj} = 1 \quad (\forall j) \\
              & \textstyle \sum_i \delta_{ix} \pi_{iy} = \sum_j \delta_{jy} \pi_{xj} \quad (\forall xy). 
\end{align*}
Swapping the max and the min, it is clear that $\lambda_k$ becomes the Lagrange multiplier of a new constraint corresponding to the moment-matching condition:
\begin{align*}
\min_{\pi_{iy}, \pi_{i0}, \pi_{xj}, \pi_{0j} \geq 0} ~
& -\sum_i \bigg[ \pi_{i0} \varepsilon_{i0} + \sum_{y} \pi_{iy} \varepsilon_{iy} \bigg]
- \sum_j \bigg[ \pi_{0j} \eta_{0j} + \sum_{x} \pi_{xj} \eta_{xj} \bigg] \\
\text{s.t.} ~ & \textstyle \pi_{i0} + \sum_y \pi_{iy} = 1 \quad (\forall i) \\
              & \textstyle \pi_{0j} + \sum_x \pi_{xj} = 1 \quad (\forall j) \\
              & \textstyle \sum_i \delta_{ix} \pi_{iy} = \sum_j \delta_{jy} \pi_{xj} \quad (\forall xy) \\
              & \textstyle \sum_{xy} \frac{1}{2} \big( \sum_i \delta_{ix} \pi_{iy} + \sum_j \delta_{jy} \pi_{xj} \big) \phi_{xyk} = \sum_{xy} \hat \mu_{xy} \phi_{xyk} \quad (\forall k).
\end{align*}
Rewriting the problem as a maximization of minus the objective yields the desired formulation.
\qed

\section{Simulations}
\label{apx:simulations}

\paragraph{Hardware.} All simulations were run on a MacBook Pro equipped with an Apple M3 Pro processor (11 CPU cores), and 18 GB RAM, running macOS Sequoia 15.2. The numerical expriments were executed using Python 3.9.13 and Gurobi 12.0.2.

\paragraph{Software.}
In our implementation of Algorithm \ref{alg:RROA} we relied on Gurobi for the iterated reoptimizations of the restricted problem \eqref{eq:restricted_problem} at each step.
We disabled the presolve options (Presolve = 0) and enabled warm starts (LPWarmStart=2).
Although we did not force Gurobi to use dual simplex for reoptimization, it consistently used it as expected by the structure of the problem. 

\begin{figure}[p]
    \centering
    \includegraphics[width=0.65\textwidth]{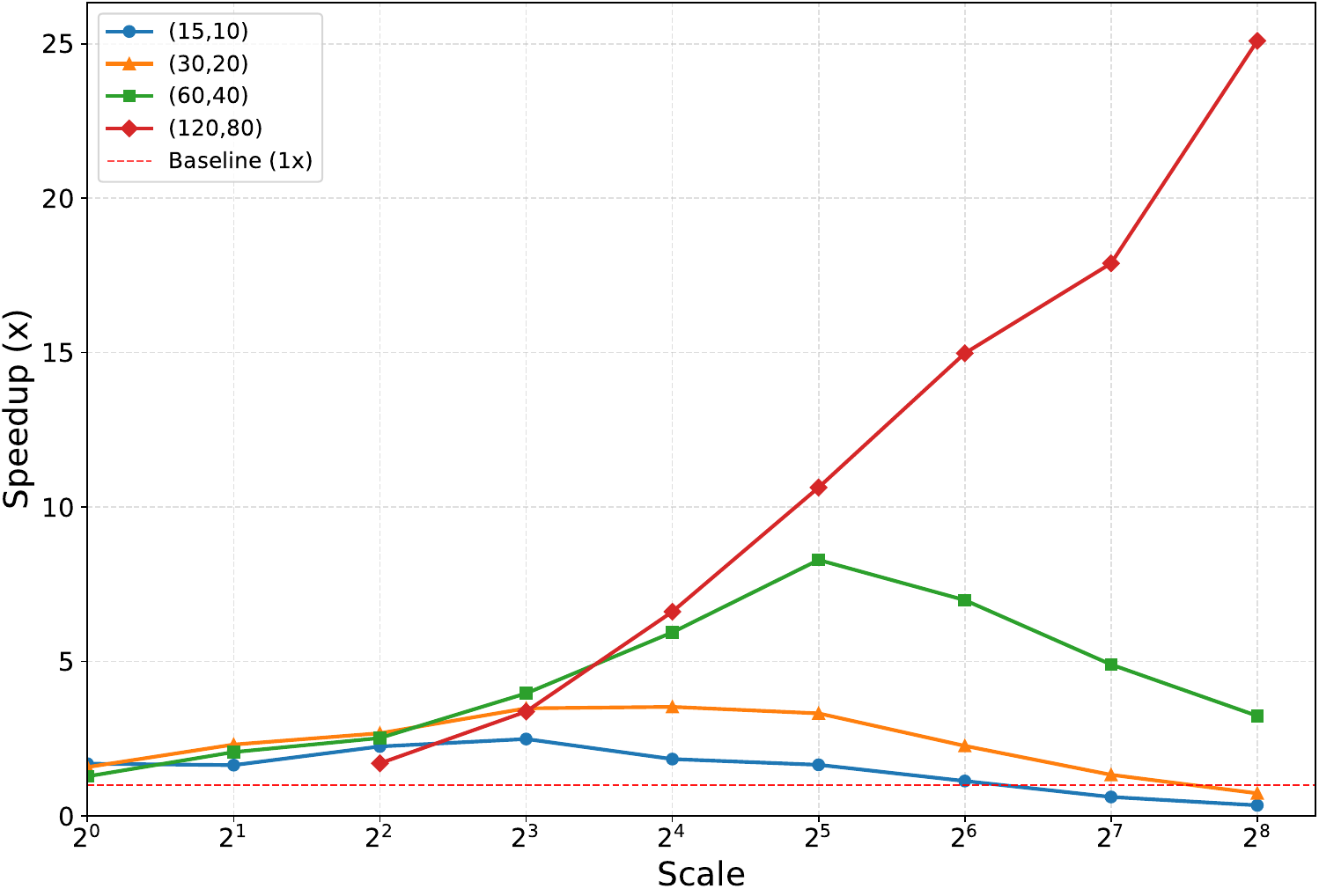}
    \caption{
    Relative performance of the RROA procedure vs.\ scale of the matching problem.
    Each line corresponds to a fixed value of the size of the type sets $(|X|,|Y|)$.
    The horizontal axis varies the size of the population using an exponential scale as $|I| = 400 \cdot S$ and $|J| = 300 \cdot S$ for $S = 1, 2, 4, \dots, 256$.
    Each point was obtained by generating and solving 10 independent assignment problems for the corresponding value of $(|X|, |Y|, S)$ and by averaging the speedup performance over those 10 problems.
    Assignment problems were generated by independently drawing surpluses $\Phi_{xy}$ from $\mathcal{N}(0, 5^2)$ and idiosyncratic preference shocks from $\mathcal{N}(0, 0.1^2)$.
    }
    \label{fig:speedup-scale-2}
\end{figure}

\paragraph{}
Figure \ref{fig:speedup-scale-2} replicates the plots of Figures \ref{fig:speedup-scale} and \ref{fig:speedup-scale-forced}, but without forcing Gurobi to use a specific method. Again, we notice that the computational gains of RROA are highly dependent on the size of the sets $X$, $Y$, $I$, and $J$.
For smaller numbers of types, the algorithm performs up to 8x faster before declining closer to 1x for larger scales. In contrast, for higher number of types, the advantage becomes substantial, and the algorithm achieves up to 25x speedups without experiencing any decline over larger population sizes. In absolute terms, this higher relative performance corresponded to computing times of around 90 seconds for our RROA method, vs. 37 minutes for Gurobi to solve one problem. 

As suggested by Figure \ref{fig:speedup-scale}, the reason behind the decline in relative performance for smaller numbers of types is due to Gurobi concurrently running several LP solvers.
For smaller population sizes, the Simplex method is faster while for greater scales (starting from around $2^4$) Gurobi begins switching to a Barrier method.
For larger sizes of the type sets, this approach also achieves up to 25x speedups, more than the 15x speedups when restricted to Barrier observed in Figure \ref{fig:speedup-scale-forced}.
The reason behind this is that when forced to use the Barrier method, Gurobi saves time by not running concurrent methods and thus solves the problem faster.
Nonetheless, in both scenarios our algorithm outperforms Gurobi in a majority of the parameter values considered.

\section{Dantzig--Wolfe}
\label{apx:Dantzig--Wolfe}

The linear programming problems \eqref{eq:optimal_matching_refined}--\eqref{eq:restricted_problem} have a special structure that makes them suitable for the Dantzig--Wolfe decomposition algorithm. 
The feasibility constraints (first $XY$ equations) link all variables across individuals, while the individual choice constraints (the last $I+J$ equations) are block-diagonal, with each individual's constraint involving only their own choice variables.

This structure allows to naturally decompose our matching problem into individual optimization problems, which are linked through the feasibility constraints. We note that this resembles the idea behind the Dantzig--Wolfe linear programming optimization algorithm and conclude that our Repeated Restricted Optimal Assignment method can be interpreted as an instance of the algorithm with the following correspondences.

\paragraph{Restricted master problem.}
In the Dantzig--Wolfe algorithm, we start with a problem that does not consider all the variables (this is known as the Restricted Master Problem). In our case, this corresponds to solving \eqref{eq:restricted_problem}. Indeed, we take the original problem \eqref{eq:optimal_matching_refined} and then restrict it using the choice sets by limiting allowed matches. Namely, an individual $i$ (resp. $j$) is only allowed to be matched with types contained in their choice set $Y_i$ (resp. $X_j$). 

This restricted formulation arises naturally from the large scale of the problem. When $I$ and $J$ are large, the full \textit{unrestricted} problem becomes computationally intractable. Instead of iterating over all possible matches, we get a much smaller linear programming problem that iteratively converges to the global optimum. This mirrors the logic of Dantzig--Wolfe.

\paragraph{Initial basic feasible solution.}
In the Dantzig--Wolfe framework, the algorithm begins with an \textit{initial basic feasible solution}, which is a solution that satisfies all the constraints. In complex large-scale problems, it is sometimes complicated to efficiently find this starting point. In our setting, this corresponds to empty choice sets, i.e.
$$Y_i^0 = X_j^0 = \emptyset$$
and
$$\pi_{i0}^0 = \pi_{0j}^0 = 1, \quad \pi_{iy}^0 = \pi_{xj}^0 = 0 \quad \forall x, y, i, j.$$  
This indeed trivially satisfies all the constraints of \eqref{eq:restricted_problem}. The individual feasibility holds since
$$\pi_{i0} + \textstyle\sum_{y \in Y_i} \pi_{iy} = 1 \quad \text{and} \quad \pi_{0j} + \textstyle\sum_{x \in X_j} \pi_{xj} = 1$$
since only $\pi_{i0}^0 = \pi_{0j}^0 = 1$.
And the linking constraints $\sum_{i : y \in Y_i} \delta_{ix} \pi_{iy} = \textstyle\sum_{j : x \in X_j} \delta_{jy} \pi_{xj}$ are clearly satisfied as both sides are just 0.
Therefore, the ``everyone is single'' initialization is a \textit{basic feasible solution} of our problem \eqref{eq:restricted_problem} and serves as a natural starting point to our algorithm, fully analogous to the initialization of Dantzig--Wolfe.

\paragraph{Reduced cost problems.}
After solving the restricted master problem, the next step is to see whether there exist new variables that can improve the objective. In the Dantzig--Wolfe setting, this question is viewed as linear programming problems (reduced cost problems), which are defined by the dual variables (shadow prices) obtained from the optimization of the restricted problem. In our setting, given the dual variables $T_{xy}, u_i, v_j$ from the optimized \eqref{eq:restricted_problem}, we get the following reduced cost problems for each individual.
For a woman $i$, the reduced cost of adding type $y$ to her choice set equals
\begin{equation}
\label{eq:women_rc_2}
\max_{y \in Y} ~ \alpha_{iy} - \sum_i \delta_{ix} T_{xy} - u_i.
\end{equation}
Similarly, for a man $j$, the reduced cost of adding type $x$ to his choice set equals:
\begin{equation}
\label{eq:men_rc_2}
\max_{x \in X} ~ \gamma_{xj} + \sum_j \delta_{jy} T_{xy} - v_j.
\end{equation}
These expressions have very clear economic interpretations: the equation \eqref{eq:women_rc_2} measures the opportunity of a woman $i$ to increase her utility. Same applies for men. When these reduced costs are positive, adding the option that was yielded from \eqref{eq:women_rc_2} (resp. \eqref{eq:men_rc_2}) increases both the individual's utility and the overall objective (surplus). The algorithm iteratively chooses the options that yield the best increase until none of the reduced costs are positive, which means that the algorithm has converged. \\
Moreover, note that, in practice, it is quite easy to compute. We can just iterate over all possible choices for each individual and this indeed will be computationally efficient as $X$ and $Y$ are practically small. For instance, suppose a person is classified by education (5 levels), ethnicity (3 groups), and age (3 categories).  This yields $5\times3\times3=45$ possible types, which is small enough to be computed by enumeration. \\
As a result, we can simply compute each reduced cost without using any linear programming solvers. This makes each iteration light, while still preserving the logic of Dantzig--Wolfe. 

\paragraph{Column generation.}
After obtaining the columns that improve the objective, the next step is to update the problem. In the Dantzig--Wolfe framework, this means adding these new columns to the restricted master and then re-optimizing it. The process is called column generation and is repeated until no subproblem yields a column that would improve the overall objective. In our algorithm, adding a new column is equivalent to allowing a new potential match in an individual's choice set. Namely, if at step $t$ of the algorithm, a subproblem for a woman $i$ (resp. man $j$) yielded a positive reduced cost $y_i^t$ (resp. $x_j^t$), then we add them to the choice set
\begin{equation*}
Y_i^{t+1} = Y_i^t \cup \{y_i^t\} \qquad (\text{resp. } X_j^{t+1} = X_j^t \cup \{x_j^t\})
\end{equation*}
or keep it the same otherwise.

\end{document}